\documentclass{osa-article}

\journal{oe}

\usepackage{siunitx}
\usepackage{placeins}
\usepackage{chemformula} 

\newcommand{\un}   [1]{\ensuremath{\,\mathrm{#1}}}

\newcommand{\neff} {\ensuremath {n_{\mathrm{eff}}}}

\articletype{Research Article}

\begin{document}

\title{Aluminum nitride integration on silicon nitride photonic circuits: a new hybrid approach towards on-chip nonlinear optics}

\author{Giulio Terrasanta\authormark{1,2}, Timo Sommer\authormark{1,4}, Manuel M{\"u}ller\authormark{3,1}, Matthias Althammer\authormark{3,1}, Rudolf Gross\authormark{3,1,4} and Menno Poot \authormark{1,4,5,*}}

\address{
\authormark{1}Department of Physics, Technical University Munich, Garching, Germany\\
\authormark{2}Physics Section, Swiss Federal Institute of Technology in Lausanne (EPFL), Switzerland\\
\authormark{3}Walther-Mei\ss ner-Institut, Bayerische Akademie der Wissenschaften, Garching, Germany\\
\authormark{4}Munich Center for Quantum Science and Technology (MCQST), Munich, Germany\\
\authormark{5}Institute for Advanced Study, Technical University Munich, Garching, Germany}

\email{\authormark{*}menno.poot@tum.de} 



\begin{abstract*}
Aluminum nitride (AlN) is an emerging material for integrated quantum photonics due to its large $\chi^{(2)}$ nonlinearity. 
Here we demonstrate the hybrid integration of AlN on silicon nitride (SiN) photonic chips. Composite microrings are fabricated by reactive DC sputtering of c-axis oriented AlN on top of pre-patterned SiN. This new approach does not require any patterning of AlN and depends only on reliable SiN nanofabrication. This simplifies the nanofabrication process drastically. Optical characteristics, such as the quality factor, propagation losses and group index, are obtained. Our hybrid resonators can have a one order of magnitude increase in quality factor after the AlN integration, with propagation losses down to \SI{0.7}{dB/cm}. Using finite-element simulations, phase matching in these waveguides is explored.
\end{abstract*}

\section{Introduction}
Recent developments in quantum technologies and quantum computing have highlighted the need for scalable, compact, and reliable platforms for quantum operations. A promising approach in this regard is using photons as qubits, and linear optical components as operators \cite{Knill2001,kok2007linear}. Photons allow light-speed transmission with low decoherence, and the qubits can be encoded in one of the many  degrees of freedom of the photon e.g. polarization or time-bin \cite{obrien2009photonic}. However, realizing such an optical quantum circuit in bulk optics is a major challenge due to the alignment of the needed optical components \cite{Zhong1460}. A promising solution is to include all components, e.g. single photon sources, quantum circuits, and detectors, in a photonic integrated circuit (PIC), which provides advantages in size, alignment complexity, and scalability. Our approach \cite{hoch2020chip,poot2016design,poot2016characterization} is to fabricate such circuits in high-stress silicon nitride (SiN), which not only is excellent in terms of optical and mechanical properties, but also flexible and reliable in its nanofabrication \cite{Liu2021}. For instance, SiN has been employed in the fabrication of low-loss waveguides with \SI{5.5}{dB/m} attenuation \cite{pfeiffer2018}, high quality factor ($Q$) microrings for the generation of low-noise frequency combs \cite{herr2012universal}, and programmable quantum circuits \cite{arrazola2021quantum}. Nevertheless, it lacks a strong second-order nonlinearity $\chi^{(2)}$, which is needed for the realization of photon-pair sources with a large separation between pump and generated photons. We note that an effective $\chi^{(2)}$ has been realized via optical poling of SiN waveguides, with an estimated value up to \SI{3.7}{pm/V} \cite{Porcel:17}. However, the drawback of this approach is that it requires a complex optical initialization process for each single-photon source.

For this reason, aluminum nitride (AlN) has been proposed as an alternative material for integrated photonics \cite{xiong2012aluminum}, with its intrinsically strong $\chi^{(2)}\approx$ \SI{4.6}{pm/V} \cite{Chen1995}. Furthermore, its wide bandgap (\SI{6.2}{eV}) allows applications from ultra-violet to infrared wavelengths \cite{Liu_OpEx_AlN_on_sapphire_UV,Jung:14}. AlN has already been employed to realize second-order nonlinear processes on a chip, with low-loss and high-quality resonators \cite{pernice2012,Guo:16}. For instance, AlN resonators can be used as a photon-pair source by spontaneous parametric down-conversion \cite{guo2017parametric}. Moreover, AlN has proven to be excellent for several other applications, such as quantum computing \cite{wan2020large} and comb generation \cite{Jung:13}. Furthermore, dispersion control in AlN microrings was demonstrated \cite{jung_OE_AlN_dispersion}. Nevertheless, the fabrication of AlN circuitry, in particular its etching, can be challenging. The presence of surface oxides gives rise to a dead time at the start of the etching process \cite{Buttari2003}, which results in slow and not easily reproducible etching rates \cite{Shah_2014}. A high inductively coupled plasma (ICP) power exceeding \SI{500}{W} is needed to achieve sidewalls with modestly steep angles such as \SI{84}{\degree} in Ref. \cite{Lu2021}. Moreover, the etching process involves highly toxic gases such as Cl$_2$ and BCl$_3$, and the etched film surface can have a significantly increased roughness after the process \cite{Zhu2004}.

In this paper, we present our AlN/SiN composite PICs fabricated with a novel approach \cite{terrasanta_MQT_AlN_on_SiN}, that consists of sputtering c-axis oriented AlN on top of pre-patterned SiN microstructures. Thin films are grown using DC reactive magnetron sputtering, which has the advantage of being low-cost and low-temperature in comparison to other AlN fabrication techniques \cite{Iqbal2018}, such as metal organic chemical vapor deposition \cite{chiang2011} and molecular beam epitaxy \cite{Dasgupta2009}. We show that the fabricated hybrid AlN/SiN microrings have excellent quality factors as high as $5\times 10^5$ and propagation losses as low as \SI{0.7}{dB/cm}. The design, fabrication, and characterization of the high-Q AlN/SiN composite microrings are reported. The benefits of these hybrid structures are evident: They do not require any patterning of AlN and thus no complex two-step etching \cite{Surya:18} of AlN and SiN is needed. Moreover, additional freedom for dispersion engineering is added.

\section{Device design and fabrication}
\label{sec:fab}
Ever since the start of integrated photonics, microring resonators are widely used, because their tight light confinement enables compact optical cavities \cite{Bogaerts2012}. 
Our current interest in such resonators not only comes from the possibility of resonantly amplifying the nonlinear optical effects of AlN, thus providing a more efficient wavelength conversion \cite{Guo:16}, but also because they allow for the characterization of the optical properties of the hybrid photonic waveguides, such as group index, bending, and propagation losses. We study a large set of microrings with varying parameters, which consists of 230 different devices on every chip. The devices all have a bus waveguide to probe the ring resonances via transmission measurements and they differ in three parameters: ring radius $R$, ring waveguide width $W$, and separation $s$ between ring and bus waveguide. The chips are divided into two blocks with devices. The first $12\times10$ block has devices with 12 different radii ranging from 10 to \SI{130}{\mathbf{\mu}m}, 10 separations between 100-\SI{800}{nm}, and all rings have a constant width $W = \SI{1.00}{\mathbf{\mu}m}$ [see also Fig. \ref{fig:figure2}(b)]. The second $11\times10$ block has devices with 11 designed ring waveguide widths ranging from 0.60 to \SI{2.10}{\mathbf{\mu}m}, 10 separations between 100 and \SI{800}{nm}, with a constant $R=\SI{80}{\mathbf{\mu}m}$ radius. We note that the design width $W$ differs from the actual waveguide width, due to a lateral etching of about \SI{130}{nm} during the waveguide-defining etch, and because of the deposited AlN. Because of these two effects, in this work the designed width and separation are given, unless stated otherwise.

\begin{figure}[htbp]
    \centering\includegraphics[width=\textwidth]{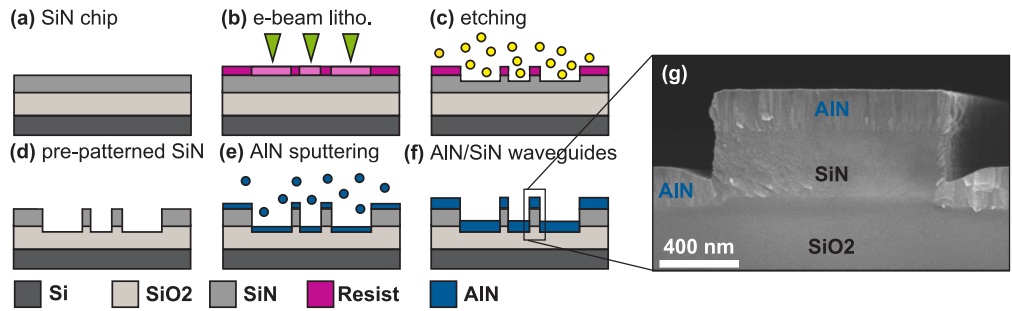}
    \caption{AlN/SiN composite nanofabrication process. (a-f) Schematic fabrication processes. (a) Starting wafer. (b) electron-beam lithography to define the PICs. (c) Etching of the waveguide. (d) Regular SiN photonic chip. (e) Growth of AlN. (f) Final hybrid PICs. (g) SEM of a waveguide cross section after growth of \SI{200}{nm} of AlN. 
    }\label{fig:figure1}
\end{figure}

As detailed in Ref. \cite{terrasanta_MQT_AlN_on_SiN}, the starting point of the hybrid fabrication is the patterning of the \SI{330}{nm}-thick SiN film. Wafers with \ch{Si3N4} on a cladding layer of \SI{3300}{nm} of SiO$_2$ on a Si substrate [Fig. \ref{fig:figure1}(a)] were purchased, and diced into smaller $6\times \SI{10}{mm}$ dies. After cleaning, ZEP520A resist (ZEON) is spin-coated and the photonic structures are defined by electron beam lithography (Nanobeam nB5) [Fig. \ref{fig:figure1}(b)]. The pattern is transferred into the SiN layer by reactive ion etching [Fig. \ref{fig:figure1}(c)] in an inductively-coupled plasma etcher (Oxford PlasmaPro 80 ICP) using an SF$_6$/CHF$_3$ chemistry. The SF$_6$ and CHF$_3$ flows, the ICP power, and the HF power are \SI{6}{sccm} and \SI{14}{sccm}, \SI{176}{W}, and \SI{7}{W}, respectively. The pattern is etched through the SiN and into SiO$_2$, with an etch depth in the SiO$_2$ layer of about \SI{50}{nm}. The patterned SiN is then covered with an AlN thin film via DC magnetron sputtering [Fig. \ref{fig:figure1}(e)]. A 3-inch pure aluminum (Al) target is sputtered in a nitrogen (N$_2$) and argon (Ar) gas mixture. The N$_2$/Ar flow, the DC power, the pressure and the substrate temperature during deposition  are respectively 3.5/\SI{10}{sccm}, \SI{210}{W}, \SI{5e-3}{mbar} and \SI{600}{\celsius}. These parameters are the result of a growth optimization series to optimize the crystalline (0002)-reflection of c-axis AlN, as detailed in Ref. \cite{terrasanta_MQT_AlN_on_SiN}. The sputtered AlN films were highly c-axis oriented: the X-ray diffraction (XRD) rocking curve of the (0002) diffraction peak had a full-width half maximum (FWHM) of \SI{3.9}{\degree} for \SI{300}{nm} thick AlN films on SiN, which is the expected range for the AlN growth on SiN and SiO$_2$ \cite{Felmetsger2011,Belkerk2014}. The surface morphology was characterized by atomic force microscopy (AFM) on a $10\times\SI{10}{\mathbf{\mu}m^2}$ area, that resulted in root-mean-square (RMS) roughness of \SI{0.47}{nm}. The refractive index, obtained using spectral reflectance, was 2.04, while the absorption coefficient was below detection limit in the visible and near-IR. Similar values were obtained for the other thicknesses. These AlN films are thus very promising for optical applications. However, the analysis in Ref. \cite{terrasanta_MQT_AlN_on_SiN} focused on the AlN films themselves and not on the hybrid photonic devices; that is the topic of this work.

To see how the AlN grows around the prefabricated waveguides, a sample with \SI{200}{nm} of AlN is diced orthogonal to the waveguides. A representative SEM image of a waveguide cross-section as shown in Fig.~\ref{fig:figure1}(g) indicates that the SiN waveguide is completely encapsulated by the sputtered AlN. The sidewalls remain steep and the AlN thickness on each sidewall is about \SI{40}{nm}, which is significantly less than the thickness on top of the waveguide. The latter is in agreement with the \SI{200}{nm} coverage far away from the waveguide, i.e. on unpatterned areas. The top part of the composite waveguide shows an overhanging `cap' shape. This indicates that not only vertical, but also horizontal growth takes place. Moreover, a `notch' is visible right next to the waveguide. From the cross section it is thus clear that near the waveguides the AlN deposition process is more complex than either the fully isotropic deposition, or the pure vertical growth that one may intuitively expect. 

\section{Results and discussion}
\subsection{Measurement setup and routine}
\begin{figure}[htbp]
\centering\includegraphics[width=\textwidth]{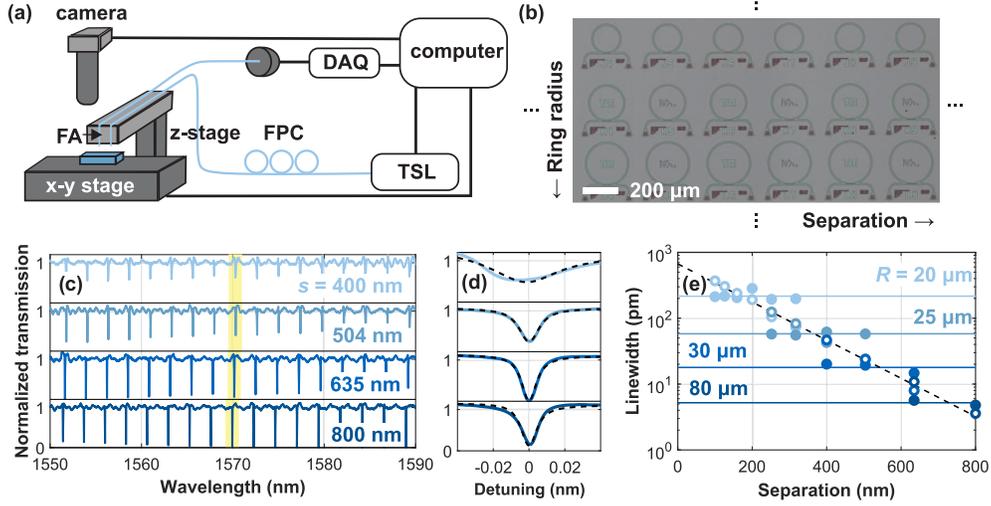}
\caption{Microring characterization procedure. (a) Schematic of the measurement setup with a tunable sweep laser (TSL), fiber polarization controller (FPC), $z$ and $x$-$y$ stages, fiber array (FA), camera, data acquisition device (DAQ) and computer. (b) Micrograph of a part of a chip, showing microring devices with \SI{31}{nm} of AlN on top. (c) Normalized transmission of hybrid devices with designed waveguide width \SI{1.00}{\mathbf{\mu}m}, ring radius \SI{80}{\mathbf{\mu}m}, AlN thickness \SI{110}{nm} and different separations listed in the plot. (d) Zooms of resonances near \SI{1570}{nm} [yellow shading in (c)] with fits (black dashed lines). (e) Internal (filled symbols/solid lines) and coupling (empty symbols/dashed lines) linewidths for listed ring radii in the plot versus microring-waveguide separation with constant design waveguide width \SI{1.00}{\mathbf{\mu}m} and an AlN thickness of \SI{110}{nm}. The linewidths are extracted and averaged in the wavelength-interval 1560-\SI{1580}{nm}. Horizontal black lines indicate the zero of each sub-panel in (c) and (d).}
\label{fig:figure2}
\end{figure}

The highly automated characterization of the optical transmission through the fabricated devices is performed  using the experimental setup shown schematically in Fig. \ref{fig:figure2}(a) \cite{hoch2020chip}. Chips are placed on a motorized $x$-$y$ stage and their position can be viewed with a microscope camera. Light is coupled onto and off the chip's waveguides using grating couplers \cite{Taillaert2002}. A tunable sweep laser (Santec TSL510) is employed as light source and the polarization of the input light is controlled by a fiber polarization controller (FPC). The FPC is connected to one of the fibers of a fiber array (FA), which is mounted on a manual $z$-stage to adjust the separation between the chip and FA. The light transmitted through the photonic devices on the chip is collected by a second fiber in the array, and guided to a photo detector (New Focus 2053-FC). The detector converts the optical power into a voltage that is finally measured with a data acquisition device (DAQ). The laser, the motorized $x$-$y$ stage, the camera and the DAQ are connected to a computer, allowing highly-automated measurements of all devices on a chip. Our automated routine locates each device, optimizes the chip position and measures the transmission reproducibly \cite{hoch2020chip}; an optical micrograph of 18 out of the 230 devices on one of the chips is shown in Fig.~\ref{fig:figure2}(b). 

\subsection{Microring characterization}
Figure \ref{fig:figure2}(c) shows normalized transmission measurements for AlN/SiN composite microrings with identical widths $W$ and radii $R$, but varying ring-bus waveguide separations $s$. The spectra still clearly show the resonances of the ring resonators, even though these now have \SI{110}{nm} AlN deposited on top of them. As shown below, their free spectral range (FSR), i.e. the spacing between the resonances, can be used to extract the group index $n_{\text{g}}$ using $n_{\text{g}}=\lambda_0^2/(2\pi R \cdot \mathrm{FSR})$, where $\lambda_0$ is the free-space wavelength \cite{Bogaerts2012}. Furthermore, the resonance linewidth can be used to obtain the propagation losses of the microring and its coupling with the bus waveguide. 

Fig.~\ref{fig:figure2}(d) shows zoom of the resonances of the normalized transmission spectra around \SI{1570}{nm}. From input-output theory, the shape of the resonance is expected to be Lorentzian \cite{Bogaerts2012} with a total linewidth that is the sum of the internal linewidth $w_{\text{int}}$ and of the coupling linewidth $w_{\text{c}}$ \cite{poot2016design}. $w_{\text{int}}$ is proportional to the losses in the resonator, such as scattering, absorption and bending losses \cite{Rath:13}, while $w_{\text{c}}$ is proportional to the coupling between the bus waveguide and the ring. The extinction of a resonance, meaning its depth, is maximal at critical coupling ($w_{\text{c}}=w_{\text{int}}$), whereas the resonator is undercoupled (overcoupled) when $w_{\text{c}} < w_{\text{int}}$ ($w_{\text{c}} > w_{\text{int}}$). Fig.~\ref{fig:figure2}(d) shows that the extinction first increases with increasing separation and then decreases again after reaching a maximum for $s=\SI{635}{nm}$. Since $w_{\text{c}}$ decreases with increasing $s$, this shows that the microring is overcoupled for a shorter separation, critically coupled around \SI{635}{nm}, and undercoupled at large $s$. This conclusion also is supported by the clearly-visible decrease in total linewidth with increasing separation. To quantify this, the resonances are fitted individually \cite{poot2016design}.
From the fits [dashed lines in Fig.~\ref{fig:figure2}(d)], the two linewidths can be extracted. Although it is a-priory not clear which one is which, the procedure from Ref. \cite{poot2016design} is used to distinguish $w_{\text{c}}$ and $w_{\text{int}}$ for each microring. 
As an example, for a subset of microrings with same width and AlN thickness but different separations and radii, $w_{\text{int}}$ and $w_{\text{c}}$ are shown in Fig. \ref{fig:figure2}(e). $w_{\text{int}}$ and $w_{\text{c}}$ are plotted with filled and open symbols, respectively. It can be observed that $w_{\text{c}}$ (open symbols) decreases exponentially with separation. The data points for different $R$ fall onto a single curve (black dashed line), meaning  that  $w_{\text{c}}$ is independent of the microring radius, at least for considered radii. Likewise, $w_{\text{int}}$ (filled symbols) is independent of separation and the averaged values (solid lines) decrease for larger ring radius, indicating a decreased bending losses for larger radii \cite{Vlasov:04}. 

\subsection{Comparison of different AlN thicknesses}
Now that the general characterization routine has been explained, the focus is set on the analysis of the device properties for different thicknesses of AlN. For this purpose, we fabricated four nominally-identical chips and deposited different AlN thicknesses (31, 74, 110, \SI{204}{nm}) as described in Sec. \ref{sec:fab}. Each chip contains 230 devices that are characterized before and after the AlN growth, thus resulting in 1840 measured and analyzed transmission spectra. Note that before AlN deposition the chips exhibit equivalent properties, and thus only one of the four chips is shown as a reference (``SiN only'') in the forthcoming analysis.

\begin{figure}[htbp]
    \centering\includegraphics[width=\textwidth]{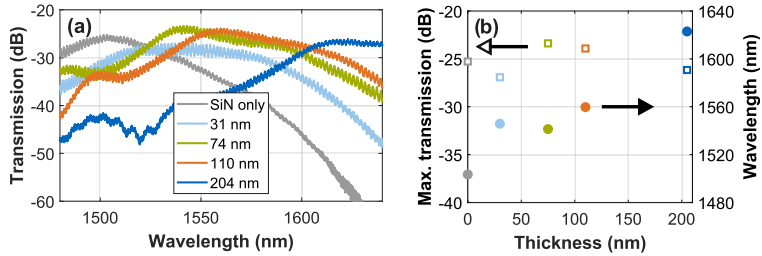}
    \caption{AlN/SiN hybrid grating coupler characterization. (a) Transmission spectra through devices without resonances with five different AlN thicknesses. (b) Extracted maximum transmission (empty symbols) and its wavelength (filled symbols) for the same spectra.}
\label{fig:figure3}
\end{figure}

First, it is discussed how the deposition of AlN affects the optical properties of the grating couplers. To study their performance for different AlN thicknesses, we measured the transmission through devices without resonances. The transmission spectra for different thicknesses of AlN are shown in Fig. \ref{fig:figure3}(a). The maximum transmission (empty symbols) and its wavelength (filled symbols) are extracted from each spectrum and plotted in Fig.~\ref{fig:figure3}(b). Without AlN, the transmission peaks at \SI{1500}{nm}; this maximum shifts to longer wavelengths when AlN is grown on top. The AlN increases the average refractive index, meaning that a longer (vacuum) wavelength is needed to match the grating period. Furthermore, Fig. \ref{fig:figure3}(a) and (b) show that after the growth of AlN, the grating couplers are still as effective as before: the maximum transmission remains at about \SI{-26}{dB}. Therefore, the employed grating coupler design is still valid for composite devices, and, if needed, wavelength shifts can be compensated by adjusting the designed period and filling factor of the grating. 

Because of the finite bandwidth of grating coupler transmission, a wavelength interval with a relatively high transmission at each AlN thickness needs to be selected for further analysis of the devices. Note that one cannot choose a different wavelength interval for each thickness, because properties such as $w_{\text{c}}$ and $w_{\text{int}}$ might differ for different wavelengths, thus compromising a fair comparison of the different AlN thicknesses. We choose the wavelength interval 1560-\SI{1580}{nm} for the rest of the paper.

\begin{figure}[htbp]
    \centering\includegraphics[width=\textwidth]{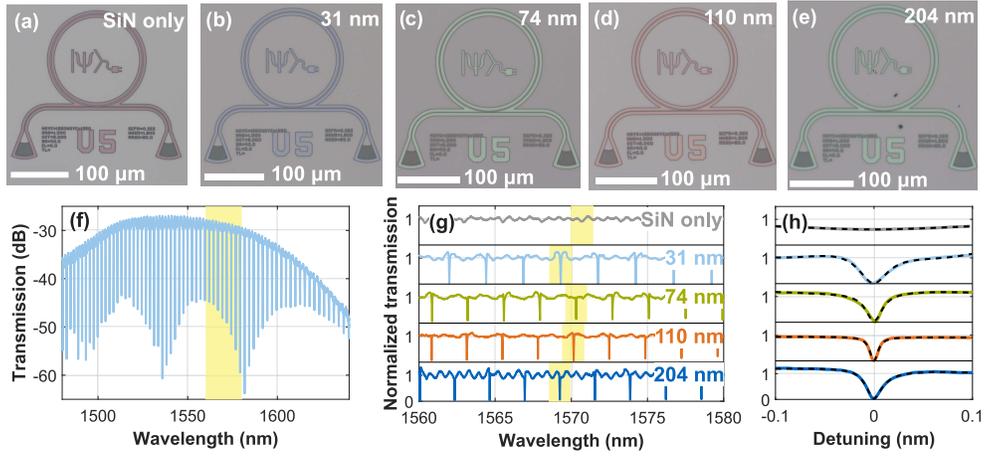}
    \caption{Comparison of microrings with different thicknesses of AlN. 
    (a)-(e) Optical images of microrings with AlN thicknesses of
    (a) \SI{0}{nm} (SiN only), 
    (b) \SI{31}{nm},  
    (c) \SI{74}{nm},  
    (d) \SI{110}{nm}, 
    (e) \SI{204}{nm}. 
    Their radius is \SI{80}{\mathbf{\mu}m}, design width \SI{1.80}{\mathbf{\mu}m} and separation \SI{252}{nm}. 
    (f) Measured transmission of a microring with $W=\SI{1.00}{\mathbf{\mu}m}$, $R=\SI{80}{\mathbf{\mu}m}$, $s= \SI{635}{nm}$ and \SI{31}{nm} AlN thickness.
    (g) Normalized transmission spectra of microring devices with different AlN thicknesses. The other parameters are as listed in caption (f). 
    (h) Zooms of resonances near \SI{1570}{nm} that were highlighted in (g) with fits (black lines).}
    \label{fig:figure4}
\end{figure}

Next, the influence of the AlN films on the microring properties are studied. Fig. \ref{fig:figure4}(a)-(e) show optical images of the same microring design in each of the considered chips. The difference in the layer thickness results in a different colour, especially visible in the etched regions. This effect underlies the reflectometry method that we use to determine the film thickness \cite{terrasanta_MQT_AlN_on_SiN}. The question is now, how these AlN films affects the resonances of the microrings.
Fig. \ref{fig:figure4}(f) shows an example of a transmission spectrum with \SI{31}{nm} of AlN. In general, every spectrum is filtered to obtain a baseline without resonances. Then, the transmission spectra are divided by these baselines to get normalized transmissions that no longer show differences due to the grating coupler transmission profiles [cf. Fig.~\ref{fig:figure3}(a)]. Fig.~ \ref{fig:figure4}(g) shows overviews of the five normalized transmissions of the same microring design with different thicknesses of AlN, whereas Fig. \ref{fig:figure4}(h) shows zooms into the resonances highlighted in yellow. The overview and zooms show that the extinction and the linewidth change with thickness, indicating a change in coupling and/or losses.

\begin{figure}[htbp]
  \centering\includegraphics[width=\textwidth]{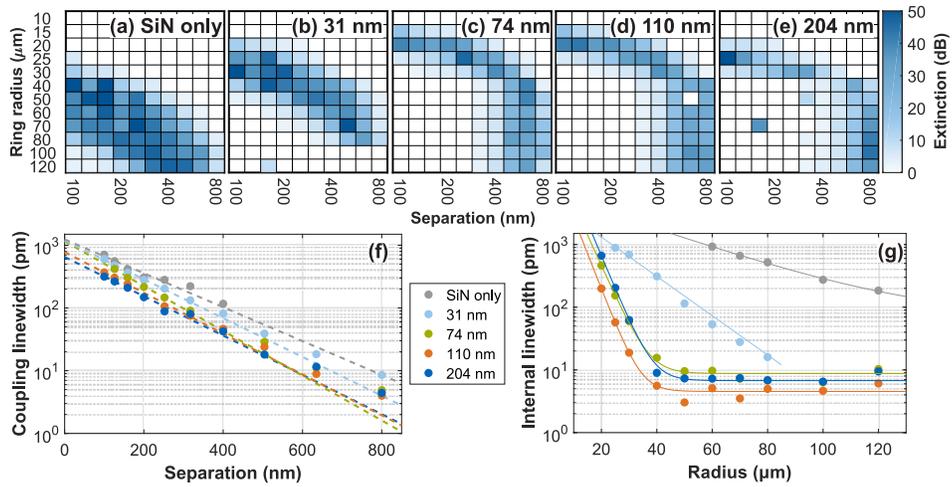}
  \caption{Extinction, $w_{\text{int}}$, and $w_{\text{c}}$ analysis in $12\times10$ block with AlN/SiN hybrid microrings with 12 different radii, 10 different separations and design waveguide width \SI{1.00}{\mathbf{\mu}m}. For this same design, five different AlN thicknesses are considered: 0 (SiN only), 31, 74, 110, \SI{204}{nm}. (a)-(e) Maximum extinction maps in the wavelength interval 1560-\SI{1580}{nm}. 
  Note that every square in (a)-(e) correspond to an individual device. The microrings in panel (b) with a radius of 100 and \SI{120}{\mathbf{\mu}m} and separation above \SI{200}{nm} could not be measured due a fabrication error.
  (f) $w_{\text{c}}$ averaged in columns where the microrings have same separation but different radii. (g) $w_{\text{int}}$ averaged in rows where the microrings have same radius but different separations.}
    \label{fig:figure5}
\end{figure}

The variations in the extinction for different microring designs provide first insights into internal losses and coupling, since the extinction depends on both $w_{\text{c}}$ and $w_{\text{int}}$ as discussed before. Fig.~\ref{fig:figure5}(a)-(e) show the maximum extinction (over the considered wavelength interval of 1560-\SI{1580}{nm}) against the ring radius and separation, as obtained for each hybrid device of the $12\times10$ block on every chip. The darker shades of blue indicate rings with higher extinction, i.e. those that are close to critical coupling. The microrings without AlN in Fig.~\ref{fig:figure5}(a) display a shift in peaked extinction towards larger separation for larger radii. From Fig. \ref{fig:figure2}(e) we know that $w_{\text{c}}$ is independent of $R$ and that it is decreasing with increased separation. Therefore, it can be concluded that $w_{\text{int}}$ is decreasing for larger radii. Therefore, bending losses dominate, even for the largest SiN-only resonators with $R=\SI{120}{\mathbf{\mu}m}$. The sample with 31 nm of AlN in Fig.~\ref{fig:figure5}(b) still shows this behavior, but with an overall shift towards the right (i.e., larger separation), indicating that the growth of AlN reduced the coupling or the propagation losses. A different trend is observed for the thicker films in Fig.~\ref{fig:figure5}(c)-(e). Here the extinction peak is no longer shifting for radii $R \gtrsim \SI{40}{\mathbf{\mu}m}$. Therefore, these AlN/SiN composite microrings are not limited anymore by bending losses. Moreover, they display smaller $w_{\text{int}}$ meaning higher quality factors and lower propagation losses. 

To quantify this result, $w_{\text{int}}$ and $w_{\text{c}}$ are obtained for every device by fitting its resonances in the wavelength interval 1560-\SI{1580}{nm}. Afterwards, $w_{\text{c}}$ is averaged over this range and also between devices with the same separation, so that a single value for $w_{\text{c}}$ is obtained for each column of Fig.~\ref{fig:figure5}(a)-(e). The result of this analysis is shown in Fig. \ref{fig:figure5}(f), where it can again be seen that $w_{\text{c}}$ is decreasing exponentially with increasing separation. An offset between the curves can also be observed. The dependence of $w_{\text{c}}$ on $s$ is typically described by an exponential function \cite{Bahadori:18}, so each set of data points is fitted with:
\begin{equation}
    w_{\text{c}}(s) = \exp \left(-\frac{s-s_{100}}{L_\text{s}} \right) \times \SI{100}{pm}.
\label{eq:wc}
\end{equation}
Here, $s_{100}$ and $L_\text{s}$ are the fitting parameters. The former, $s_{100}$, represents the separation where $w_{\text{c}}=\SI{100}{pm}$, and it useful to asses whether the waveguides need to be closer or further apart after AlN deposition in order to still have the same $\SI{100}{pm}$ coupling linewidth as before; it thus describes horizontal shifts of the fit function \cite{footnote:fit}. The latter, $L_\text{s}$, is a length scale that determines the slope of the exponential decrease; a small $L_\text{s}$ indicates a rapid change of the coupling linewidth with separation, whereas a large value indicates that $w_c$ is not very sensitive to $s$.
The fitted curves are shown as dashed lines in Fig.~\ref{fig:figure5}(f) and the fit results are summarized in Table \ref{tab:fit_result}. First, it can be observed that $s_{100}$ decreases with increased AlN thickness indicating that the bus waveguide has to be closer to the ring to keep the coupling constant. 
This shows that the stronger confinement of the optical mode for thicker AlN dominates over the reduction of the separation by the deposition on sidewalls as visible in Fig. \ref{fig:figure1}(g).
The fit result for $L_\text{s}$ do not show a clear trend; the AlN thus does not seem to have a strong effect on the length scale of the coupling. Overall, since $w_\text{c}$ closely follows Eq. \eqref{eq:wc}, it can be concluded that the coupling in the AlN/SiN composite structures can be controlled by adjusting $s$ for a given AlN thickness to obtain the targeted post-deposition $w_c$. 

\begin{table}[htbp]
  \centering
  \caption{Results of fitting Eqs. \eqref{eq:wc} and \eqref{eq:wint} to the obtained coupling linewidth $w_{\text{c}}$ (center) and internal linewidth $w_{\text{int}}$ (right), respectively. The fitted curves of $w_{\text{c}}$ and $w_{\text{int}}$ were shown as lines in Fig.~\ref{fig:figure5}(f) and  Fig.~\ref{fig:figure5}(g), respectively. The uncertainty in the fit parameters is the $1\sigma$ value of the fit.}
    \begin{tabular}{c|cc|ccc}
    \hline
    AlN thick. (nm) & $s_{\text{100}}$ (nm) & $L_{\text{s}}$ (nm) & $R_{\text{100}}$ ($\mu$m) & $L_{\text{R}}$ ($\mu$m) & $w_{\infty}$ (pm) \\
    \hline
    0     & 405$\pm$39  & 161$\pm$24 &  122$\pm$20  &  29$\pm$8  & - \\
    31    & 348$\pm$9 & 140$\pm$6 & 56$\pm$6 &  14$\pm$3  & - \\
    74    & 298$\pm$6 &  122$\pm$5 &  26.72$\pm$0.11 & 4.43$\pm$0.07  & 8.7$\pm$1.0 \\
    110   &  274$\pm$5 & 134$\pm$5  & 22.54$\pm$0.05 & 3.84$\pm$0.06  & 4.5$\pm$0.5 \\
    204   & 260$\pm$7 &  139$\pm$7 &  27.75$\pm$0.07 & 4.12$\pm$0.04  & 6.7$\pm$0.8 \\
    \hline
    \end{tabular}%
  \label{tab:fit_result}%
\end{table}%

A similar analysis is performed for $w_{\text{int}}$, which is averaged for the same radius (one averaged $w_{\text{int}}$ value for each row); the results are displayed in Fig.~\ref{fig:figure5}(g). It can be observed that the sets of points with SiN only and \SI{31}{nm} of AlN always decrease for increased radius. Instead, the sets of points for the other three thicknesses (74, 110, \SI{204}{nm}) display a saturation for $R\gtrsim\SI{40}{\mathbf{\mu}m}$. This is in agreement with Fig. \ref{fig:figure5}(a)-(e), where it was already concluded that $w_{\text{int}}$ of both \SI{0}{nm} and \SI{31}{nm} AlN was always dominated by bending losses. Nevertheless, there is already a significant reduction of the propagation losses, since $w_{\text{int}}(31)$ is at least one order of magnitude smaller than $w_{\text{int}}(0)$ for all $R$ (the number in parentheses indicates the AlN thickness in nm). Even thicker AlN layers do not suffer from bending losses for $R\gtrsim\SI{40}{\mathbf{\mu}m}$. The points are fitted using
\begin{equation}
  w_{\text{int}}(R) = w_{\infty} + \exp\left(-\frac{R-R_{100}}{L_\text{R}}\right) \times \SI{100}{pm},
\label{eq:wint}
\end{equation}
where $R$ is the microring radius, and $w_{\infty}$, $R_{100}$, $L_\text{R}$ are fitting parameters. $w_{\infty}$ represents the non-bending internal losses, e.g. those due to absorption and scattering, and quantifies the saturation $w_{\text{int}}(R\rightarrow \infty)$ observed for large radii. The right part of Eq. \eqref{eq:wint} reflects the bending losses which in the simplest approach can be described with an exponential function \cite{Vlasov:04}. The parameter $R_{100}$ represents the radius needed to have bending losses equivalent to \SI{100}{pm}, whereas as before $L_\text{R}$ determines the slope of the exponential. 
The solid lines in Fig. \ref{fig:figure5}(g) show that Eq.~\eqref{eq:wint} fits the data well and the fit results are summarized in Table \ref{tab:fit_result}. $R_{100}$ is smaller after the integration of AlN, quantifying the reduced bending losses of the hybrid AlN/SiN microrings visible in Fig. \ref{fig:figure5}(g). A similar trend can also be observed for $L_\text{R}$: its values are lower after the AlN deposition, meaning that the bending losses decrease faster with increasing radius. This is also visible by the steeper slopes in Fig.~\ref{fig:figure5}(g). The fit results for the propagation loss without the bending loss contribution, $w_{\infty}$, show values below \SI{10}{pm} for the thicker films; note that the internal losses for SiN-only microrings and for \SI{31}{nm} of AlN could not be obtained as that would require larger $R$. Still, the data in Fig.~\ref{fig:figure5}(g) suggests that also $w_{\infty}(31) < 10 \un{pm}$. 
Thus, the integration of AlN on the SiN microring still exhibits small internal losses, proving the good quality of our AlN/SiN composite structures. For a quantitative analysis, the propagation losses can be calculated from $w_{\text{int}}$. For this, and for comparison with literature, the internal linewidth is first converted to the quality factor $Q_{\text{int}} \equiv \lambda _0 / w_{\text{int}}$. The best quality factor is obtained for $w_{\text{int}}(110)=3.03\pm$\SI{0.15}{pm} at radius \SI{50}{\mathbf{\mu}m} [Fig. \ref{fig:figure5}(g)], which results in a quality factor $Q_{\text{int}} = (5.19\pm0.26)\times 10^5$. This value is comparable to $Q_{\text{int}} \approx 4\times 10^5$ of composite AlN/SiN microrings \cite{Surya:18}, and to AlN-only resonators with total $Q \approx 2\times 10^5$ \cite{pernice2012} in the weakly coupled regime and $Q_{\text{int}} \approx 4.60\times 10^5$ \cite{Guo:16}. Using $n_{\text{g}}=2.125$ (see below) and $\lambda=$ \SI{1570}{nm}, the resulting propagation loss is \cite{Rath:13} 
$\alpha=(2\pi n_{\text{g}})/(Q_{\text{int}}\lambda)$= \SI{0.16}{cm^{-1}}, which corresponds to $\alpha_{\text{dB}}=10\log_{10}(\text{e})~\alpha=$ \SI{0.71}{dB~cm^{-1}}. This value is comparable with the \SI{0.6}{dB~cm^{-1}} propagation losses of an AlN-only waveguide reported in Ref. \cite{xiong2012aluminum}.

\begin{figure}[htbp]
  \centering\includegraphics[width=\textwidth]{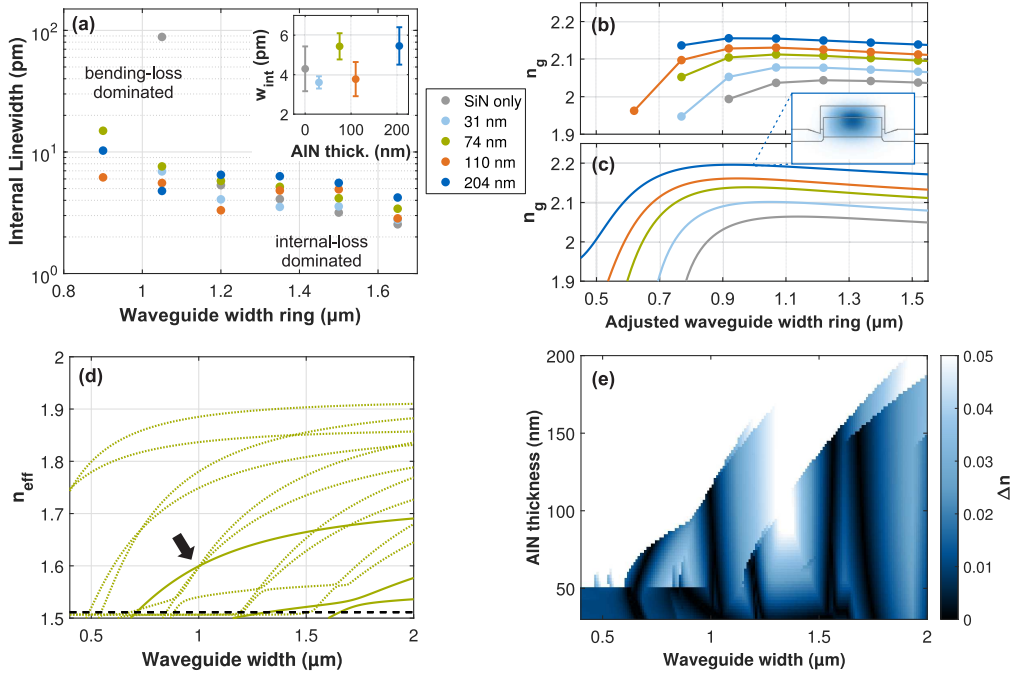}
  \caption{
  Width and thickness dependence of $w_{\text{int}}$ and $n_{\text{g}}$ for $R=\SI{80}{\mathbf{\mu}m}$. 
  (a) $w_{\text{int}}$ averaged in rows where the microrings have the same waveguide width, but a different separation. The inset shows the the average and standard deviation of the last four points. 
  (b) Experimental $n_{\text{g}}$ averaged in the same rows. For comparison with the simulations in (c), the lateral etch distance of 130 nm is subtracted from the designed width. 
  (c) Simulated $n_{\text{g}}$ using COMSOL Multiphysics\textsuperscript{\textregistered}. The inset shows the extracted mode profile from the simulation.
  (d) Simulated waveguide-width dependence of $\neff$ for a fixed  AlN thickness of $74 \un{nm}$ for both $\lambda_0 = 1550$ (solid line) and 775 nm (dotted). The black line indicates the effective index of the first slab mode \cite{saleh_teich} for $\lambda_0 = 775 \un{nm}$. Phase matching is indicated by the arrow.
  (e) Color plot of the mismatch $\Delta n$.}
  \label{fig:figure6}
\end{figure}

To realize nonlinear optical processes such as second-harmonic generation (SHG) and degenerate spontaneous parametric down conversion (SPDC), the phase-matching condition between light with vacuum wavelength $\lambda_0$ and its second harmonic at $\lambda_0/2$ has to be satisfied: $\neff(\lambda_0)=\neff(\lambda_0/2)$ \cite{pernice2012}. The effective refractive index of the guided optical mode $n_{\text{eff}}$ can be controlled by changing the geometry of the waveguide, typically its width and height. Since the (pre-growth) height is determined by the SiN wafer, we focus on studying the properties of microrings with different waveguide width $W$. To do so, the chip design also contains an $11\times10$ block of microrings with 10 different separations and 11 different widths. 
The analysis of this block is performed with a similar method to the previous one where $R$ varied. Here $w_{\text{int}}$ is obtained for each waveguide width $W$ and the results are plotted in Fig. \ref{fig:figure6}(a). The resonators without AlN (gray) suffer high losses for $W<$\SI{1.20}{\mathbf{\mu}\m}. That is in agreement with the bending losses-limited $w_{\text{int}}$ of Fig. \ref{fig:figure5}(g), where $W = 1.00 \un{\mu m}$ was used. After the AlN deposition, those losses were reduced by more than one order of magnitude. At $W = 1.05 \un{\mu m}$, there is still a large reduction of the bending loss, but this drastic improvement is no longer visible for larger widths $W\ge 1.20 \un{\mu m}$, where the microrings are not suffering from bending losses, but are limited by other internal losses such as scattering and absorption. Here, the impact of the AlN on these loss mechanisms can be studied. First of all, we note that the ordering of the points in Fig.~\ref{fig:figure6}(a) does not stay the same for the last four widths and that their values are similar, indicating no strong width dependence for $W\ge 1.20 \un{\mu m}$. The inset of Fig \ref{fig:figure6}(a) shows the last four points of Fig. \ref{fig:figure6}(a) averaged for each AlN thickness. Also here, the points do not show a clear trend with the thickness of the AlN film and agree with the intrinsic loss of the SiN-only sample. Therefore, it can be concluded that the deposited AlN does not have a strong influence on the absorption and scattering. 

As explained above, for phase matching, it is the refractive index that is important. The group index $n_{\text{g}}$ calculated from the FSR is extracted for each microring and averaged for rings with varying $s$ but identical waveguide widths. Fig.~\ref{fig:figure6}(b) shows $n_{\text{g}}$ as a function of the ring waveguide width and of the AlN thickness. For the horizontal axis of this plot, the design width was decreased by \SI{130}{nm} to account for lateral etching. This value was obtained from measurements of the waveguides cross section, such as the one shown in Fig.~\ref{fig:figure1}(g).
$n_{\text{g}}$ is also simulated using finite-element simulations and their results are displayed in Fig.~\ref{fig:figure6}(c). In both cases, $n_{\text{g}}$ increases for a thicker AlN layer. Furthermore, the overall shape of the experimental group index is also similar to the simulation: the shift of the maximum in $n_\text{g}$ with increasing thickness, and the shifted cut-off (visible as the sharp decrease in $n_\text{g}$ for narrower waveguides) are visible in both the experimental data as well as in the simulations. A closer inspection, however, reveals that there is a $\sim 0.05$ difference in $n_{\text{g}}$ between the curves, which may be due to small differences between the tabulated refractive index of AlN \cite{pastrnak_PSSB_AlN_refindex} and the actual value of our film. Still, with these simulations, also the phase matching can be explored. Figure~\ref{fig:figure6}(d) shows the calculated effective index for light propagating through hybrid waveguides for $\lambda_0 = 1550$ (solid line) and for its second harmonic at 775 nm (dotted). Near a waveguide width of $1.0 \un{\mu m}$, the lines cross. In other words, the phase matching condition is fulfilled (black arrow). The modes involved are the fundamental TE$_{00}$-like, and the  TE$_{20}$-like and TM$_{20}$-like modes at 775 nm. These simulations demonstrate that phase matching in these hybrid waveguides is possible. However, these now have an additional degree of freedom, namely the AlN thickness. Figure \ref{fig:figure6}(e) displays the smallest mismatch $\Delta n = \min_{i,j} | n_{\textrm{eff},i}(1550 \un{nm}) - n_{\textrm{eff},j}(775 \un{nm})|$ between the $\neff$ of the fundamental modes $i$ and the second harmonic modes $j$ \cite{footnote:deltan}. A large $\Delta n$ (light colors) indicates a large phase mismatch, whereas for $\Delta n = 0$ (black) phase matching is achieved. The color plot shows several dark-colored regions where this is indeed the case. Interestingly, there are dark regions that are almost vertical where the phase matching condition is, thus, insensitive to variations in the film thickness, as well as regions with a finite slope. In the former, the phase matching will be robust against variations in the deposition process \cite{terrasanta_MQT_AlN_on_SiN}, whereas for the latter the AlN thickness provides an additional degree of freedom to control phase matching. This thus shows the potential of our hybrid AlN/SiN waveguides for nonlinear optic applications, such as second-harmonic generation and spontaneous parametric down conversion.
\FloatBarrier

\section{Conclusion and outlook}
Hybrid AlN/SiN microring resonators were fabricated by DC reactive magnetron sputtering of highly c-axis oriented AlN on top of pre-patterned SiN photonic circuits. This novel hybrid fabrication approach does not require any patterning of AlN, thus drastically simplifying the nanofabrication. The hybrid structures have the advantage of combining the flexible fabrication of SiN with the nonlinear properties of AlN, which can be a promising solution to achieve efficient second-order nonlinear processes on a SiN chip. 

Microring resonators with different radii, separations to the bus waveguide, waveguide widths, and AlN thicknesses were characterized with a highly automated measurement routine. First, the hybrid grating couplers were discussed, and it was shown that the AlN deposition does not decrease the maximum transmission, but shifts the maximum peak to longer wavelengths. Afterwards, the internal linewidth $w_{\text{int}}$ and the coupling linewidth $w_{\text{c}}$ of the microrings were extracted and studied to analyze losses and coupling as a function of the AlN thickness. The losses in the resonators are decreased after the growth of AlN due to a reduction in bending losses, resulting in quality factors up to $5\times 10^5$. Moreover, the group index was extracted and showed values in agreement to simulations, which were then used to explore phase matching of light at 1550 nm and its second harmonic. Regions with phase matching were found and it was shown that the AlN-thickness can be used as an additional degree of freedom.

The next step is to test the nonlinear properties of the hybrid waveguides, by fabricating phase-matched microring resonators and to measure second harmonic generation as well as the inverse process where single photons are generated using spontaneous parametric down conversion \cite{Guo:16}. 

\section*{Acknowledgments}
The authors acknowledge support of David Hoch and Xiong Yao with the nanofabrication of the photonic circuitry. 
This project is funded by the German Research Foundation (DFG) under Germany's Excellence Strategy - EXC-2111 - 390814868 and by TUM-IAS, which is funded by the German Excellence Initiative and the European Union Seventh Framework Programme under grant agreement 291763. 

\section*{Disclosures}
The authors declare no conflicts of interest.

\section*{Data Availability Statement}
The data that support the findings of this study are available upon reasonable request from the authors.

\bibliography{AlNrings}

\begin{thebibliography}{10}
\bibitem{footnote:fit}
We note that Eq.~\eqref{eq:wc} can also be written as $w_\text{c} =
  w_{\text{c},0} \exp(-s/L_s)$. However, we find the fit parameter
  $w_{\text{c},0} = \exp (s_{100}/L_\text{s}) \times \SI{100}{pm}$ that
  quantifies the coupling linewidth at $s = 0$, less straightforward to
  interpret than $s_{100}$, which refers directly to a physical dimension of
  the device, even although there may be some arbitrariness in our choice of
  \SI{100}{pm}.
\bibitem{footnote:deltan}
The minimization of $\Delta n$ over the modes $i$ and $j$ is done with modes
  $j$ satisfying $n_{\mathrm{eff},j}(775 \un{nm}) > n_\mathrm{slab}(775
  \un{nm})$ only. Modes with effective indices below that of the first slab
  mode, $n_\mathrm{slab}(775 \un{nm})$, are typically not located at the
  waveguide, but instead in the AlN slab. The overlap of these modes with
  waveguide modes is, thus, small and SHG will not be generated efficiently.
  The jumps visible in \ref{fig:figure6}(e) are due to the (dis)appearance of
  one of the modes $j$ from the minimization when it crosses the effective
  index of the slab mode.
\end{thebibliography}

\end{document}